\documentclass[12pt]{article}
\usepackage{graphicx,amsmath}
\usepackage{color}

\newlength{\dinwidth}
\newlength{\dinmargin}
\def\be{\begin{equation}}   
\def\ee{\end{equation}}  
\def\bea{\begin{eqnarray}}                      
\def\eea{\end{eqnarray}}
\def\ch1{$\chi(1^+)$}
\def\lapproxeq{\lower .7ex\hbox{$\;\stackrel{\textstyle                                                    
<}{\sim}\;$}}                                                    
\def\gapproxeq{\lower .7ex\hbox{$\;\stackrel{\textstyle                                                    
>}{\sim}\;$}}

\begin{document}

\begin{flushright}                                                    
IPPP/17/9\\
\today \\                                                    
\end{flushright} 

\vspace*{0.5cm}

\vspace*{0.5cm}
\begin{center}
{\Large \bf Optimal choice of factorization scales for the } \\ 
\vspace{0.2cm}
{\Large \bf description of jet production at the LHC}\\
\vspace*{1cm}

A.D. Martin$^{a}$ and M.G. Ryskin$^{a,b}$\\

\vspace*{0.5cm}
$^a$ Institute for Particle Physics Phenomenology, Durham University, 
Durham, DH1 3LE, UK\\ 

$^b$ Petersburg Nuclear Physics Institute, NRC `Kurchatov Institute', 
Gatchina, St.~Petersburg, 188300, Russia \\
\end{center}

\vspace{1cm}

\begin{abstract}
 To obtain more precise parton distribution functions (PDFs) it is important to include data on inclusive high transverse energy jet production in the global parton analyses.
These data have high statistics and the NNLO terms in the perturbative QCD (pQCD) description are now available.
Our aim is to reduce the uncertainty in the comparison of the jet data with pQCD.  To ensure the best convergence of the pQCD series it is important to choose 
the appropriate factorization scales, $\mu_F$. We show that it is possible to absorb and resum in the incoming PDFs
and fragmentation function ($D$) an essential part of the higher 
$\alpha_s$ order corrections by determining the `optimal' values of $\mu_F$.
We emphasize that it is necessary
 to optimize different factorization scales for the various factors in the cross section: indeed, both of the PDFs, and also the fragmentation function, have their own optimal scale.  We show how the values of these scales can be calculated for the LO (NLO) part of the pQCD prediction of the cross section based on the theoretically known NLO (NNLO) corrections. After these scales are fixed at their optimal values, the residual factorization scale dependence is much reduced.

\end{abstract}
\vspace*{0.5cm}

\section{Introduction} 
With the availability of the complete QCD formulation of jet production to NNLO \cite{CGP} we are entering the precision era for extracting parton distribution functions (PDFs) from including these data \cite{ATLAS,CMS} in the global PDF analyses.  However, we have to address the problem of the optimal choice of factorization scales.  Here there are two problems.  The first concerns the definition of a jet. In particular, the vector sum, $p_{T\rm jet}$, of the transverse momenta of particles measured inside a jet cone $\Delta R$ is not equal to the transverse momentum $p_T$ of the parton. In fact, for large $\Delta R$ we may have  $p_{T\rm jet}>p_T$ when two large $p_T$ partons occur in the jet cone, while for small $\Delta R$ we have $p_{T\rm jet}<p_T$ since, due to final parton showering, part of the energy is emitted outside the jet cone. The second problem is how to choose the factorization scales for the production of multi-particle systems  which minimize the next fixed-order perturbative QCD (pQCD) correction to this process.

\section{The origin of factorization scales}

From a formal point of view, factorization scales are unphysical quantities.  The final result should not depend on their choice. They are introduced into pQCD just for convenience to separate the part of the cross section described by the hard matrix element for the partonic subprocess of interest from the part that can be described by PDFs or fragmentation functions which are {\it universal} and do not depend on the particular subprocess. Depending on the choice of factorization scales, a larger or smaller part of a fixed-order contribution is placed in the matrix element.  As a rule, it is advantageous to move a major part of the higher-order corrections into the universal PDFs and to minimize the remaining contribution in the matrix element.

\subsection{An example}

It is useful to illustrate the procedure in terms of a simple example. Therefore before discussing jet production, let us first consider open $b\bar{b}$ production \cite{OMR}.

The cross section for open $b\bar{b}$ production at LO + NLO calculated with factorization scale $\mu_f$ may be expressed in the form\footnote{For
  ease of understanding we omit the parton labels $a=g,q$ on the
  quantities in~(\ref{eq:sta}) and the following equations. The matrix
  form of the equations is implied.} 
\begin{equation}
\sigma^{(0)}(\mu_f) + \sigma^{(1)}(\mu_f) ~=~ \alpha_s^2\left[{\rm PDF}(\mu_f)\otimes C^{(0)} \otimes {\rm PDF}(\mu_f) +
{\rm PDF}(\mu_f)\otimes\alpha_s C^{(1)}(\mu_f) \otimes {\rm PDF}(\mu_f)\right]\,, 
\label{eq:sta}
\end{equation}
where the coefficient function
$C^{(0)}$ does not depend on the factorisation scale, while the $\mu_f$ dependence of the NLO coefficient function arises since we have to subtract from the NLO diagrams the  part already generated by LO evolution. 

We are
free to evaluate the LO contribution at a different scale $\mu_F$,
since the resulting effect can be {\it compensated} by changes in the
NLO coefficient function, which then also becomes dependent on
$\mu_F$. In this way eq.~(\ref{eq:sta}) becomes 
\begin{equation}
\sigma^{(0)}(\mu_f) + \sigma^{(1)}(\mu_f) ~=~\alpha_s^2\left[{\rm PDF}(\mu_F)\otimes C^{(0)} \otimes {\rm PDF}(\mu_F) +
{\rm PDF}(\mu_f)\otimes\alpha_s C_{\rm rem}^{(1)}(\mu_F) \otimes {\rm PDF}(\mu_f)\right]\,.  
\label{eq:stab1}
\end{equation}
Here the first $\alpha_s$ correction $C^{(1)}_{\rm rem}(\mu_F)\equiv C^{(1)}(\mu_f=\mu_F)$ is now calculated  at the scale $\mu_F$ used for the LO term, and not at the scale $\mu_f$ corresponding to the cross section on the left hand side of the formula. Since it is the correction which {\em remains} after the factorization scale in the LO part is fixed, we denote it by $C_{\rm rem}^{(1)}(\mu_F)$.
Note that although the first and second terms on the right hand side
depend on $\mu_F$, their sum, however, does not (to ${\cal O}(\alpha_s^4)$), and
is equal to the full LO+NLO cross section calculated at the factorization
scale $\mu_f$. 

Originally the NLO coefficient functions $C^{(1)}$ are calculated
from Feynman diagrams which are independent of the factorization
scale. How does the $\mu_F$ dependence of $C^{(1)}_{\rm rem}$
in~(\ref{eq:stab1}) actually arise? It occurs because we must subtract 
from $C^{(1)}$ the $\alpha_s$ term which was already included in the
LO contribution. Since the LO contribution
was calculated up to some scale $\mu_F$ the value of $C^{(1)}$ after
the subtraction depends on the value $\mu_F$ chosen for the LO
component. The change of scale of the LO contribution from $\mu_f$ to
$\mu_F$ also means we have had to change the factorisation scale which
enters the coefficient function $C^{(1)}$ from $\mu_f$ to $\mu_F$. 
Moreover, we are allowed to use different scales $\mu_f=\mu_-$ and $\mu_f=\mu_+$ for the left and right PDFs respectively.
The
effect of these scale changes is driven by the LO DGLAP evolution, which
is given by
\be
\sigma^{(0)}(\mu_F)~=~
\alpha_s^2~{\rm PDF}(\mu_-)\otimes \left(C^{(0)} +\frac{\alpha_s}{2\pi}\left[
    \ln\left(\frac{\mu_F^2}{\mu_-^2}\right) P_{\rm left}\otimes C^{0)} +\ln\left(\frac{\mu_F^2}{\mu_+^2}\right)C^{(0)} \otimes P_{\rm right}\right]\right)
\otimes {\rm PDF}(\mu_+)\,,
\label{eq:dg}
\ee
where $P_{\rm left}$ and $P_{\rm right}$ denote the DGLAP splitting functions acting on the PDFs to the left and right respectively. That is, by choosing
to evaluate $\sigma^{(0)}$ at scale $\mu_F$ we have moved the part of the NLO
(i.e. $\alpha_s$) correction given by the ${\cal O}(\alpha_s^3)$ terms of~(\ref{eq:dg})
from the NLO to the LO part of the cross section. In this way $C^{(1)}$
becomes the remaining $\mu_F$-dependent coefficient function
$C^{(1)}_{\rm rem}(\mu_F)$ of~(\ref{eq:stab1}).
The idea for open $b\bar{b}$ production at {\it low} $x$ was to choose a scale $\mu_F=\mu_0$ such that the remaining NLO term does not contain the important double-logarithmic $\alpha_s{\rm ln}( \mu_F){\rm ln}(1/x)$  contribution; in fact all the $(\alpha_s{\rm ln}( \mu_F){\rm ln}(1/x))^n$ are resummed in the PDFs. 

In this low $x$ limit, we may neglect the ${\cal O}(x)$ power corrections and the situation becomes left-right symmetric. Thus we have $\mu_+=\mu_-=\mu_F$.  However, in general, and in particular for high-$p_T$ forward jet production, the behaviour of the `left' and `right' PDFs are quite different. For example, one incoming parton be mainly a gluon and the other may be a valence quark. That is why we reserve the possibility to have $\mu_+\ne \mu_-$ in (\ref{eq:dg}). 

Although the discussion of open $b\bar{b}$ production has been carried out to NLO, it is possible to extend this procedure to higher orders, see for example eq. (6) of Ref.~\cite{OMR}.

\subsection{Physical understanding of the example}

In principle we may choose arbitrary factorization scales $\mu_-$ and $\mu_+$ for the incoming PDFs in (\ref{eq:dg}), accounting for the remaining contribution in the NLO matrix element (coefficient function).  Recall, however, that the logarithmic integration $\int d{\rm ln}k^2$ over the incoming parton virtuality $k^2$, hidden in the DGLAP evolution of the PDFs, does not extend up to infinity.  It is limited by the exact form of the  {\it off-shell} ($k^2_T\ne 0)$ LO matrix element ${\cal M}^{\rm LO}(k^2)$, which ensures that the integral is convergent.  The best choice of the factorization scales $\mu_\pm$ is such that the value of the logarithmic DGLAP integral up to $\mu_\pm$ is equal to the value of the respective convergent integral 
\be
\int^\infty_{Q_0^2}\frac{dk^2}{k^2}|{\cal M}^{\rm LO}(k^2)|^2~=~\int^{\mu^2_{\pm}}_{Q_0^2}\frac{dk^2}{k^2}|{\cal M}^{\rm LO}(k^2=0)|^2 ~=~|{\cal M}^{\rm LO}(k^2=0)|^2~{\rm ln}\frac{\mu_\pm^2}{Q_0^2}.
\label{eq:i}
\ee 
It means that we have moved to the LO PDF all the part of the NLO correction which has the same structure as that of the DGLAP evolution. This choice of scale is the best that we can do. Of course, there are completely different NLO contributions which cannot be reproduced by the evolution. For instance we cannot move the NLO vertex correction to the LO PDF as it will not be reproduced by the evolution. 

In order to calculate the value of the scale $\mu_{\pm}$ we need to know the $k^2$ dependence of the off-shell hard matrix element, ${\cal M}^{\rm LO}(k^2)$.  Formally, it would appear to be best to calculate ${\cal M}(k^2)$ with one incoming parton off-shell. However, such a quantity is not gauge invariant.  An alternative possibility is to use the axial gauge which provides a factorized ladder structure of PDF evolution which generates this off-shell incoming parton.  Another possibility is to consider the NLO subprocess where this parton is produced by a new on-mass-shell parton.  For open $b\bar{b}$ production from incoming gluons, we may consider the NLO $qg\to qb\bar{b}$ subprocess, where the light quark produces the off-shell gluon.

\section{Factorization scales for jet production}

We now return to inclusive jet production. As seen from the $b\bar{b}$ example, the value of the optimal scale, which minimizes the size of the next $\alpha_s$ correction, is driven by the properties of the previous order $\alpha_s$ matrix element. We see from eq. (\ref{eq:dg}) that we may choose the scales\footnote{For low $x$ $b\bar{b}$ production we actually have only one scale $\mu_+=\mu_-=\mu_0$. It was found \cite{OMR} that $\mu_0 \simeq 0.85\sqrt{p^2_T+m_b^2}$.} of the LO contribution which provides the most precise LO description of the process; that is, which have the smallest NLO correction. In NNLO jet production the aim is to choose scales which provide the best accuracy of the NLO result. That is we choose scales which move the largest possible part of the NNLO correction into the PDFs and fragmentation function. These corrections (including higher-order $\alpha_s$ contributions) will then be resummed via DGLAP evolution.

\subsection{Three scales}
For jet production we have to account for the final parton showering. That is, we have to introduce a fragmentation function $D(z,\mu_D)$. Again, in general, the factorization scale $\mu_D$ may be chosen to be  different to the other scales. Moreover, we can use different values of $\mu_D$ at each $\alpha_s$ order;
each time making the corresponding subtraction in the higher-order terms, which will now depend on the values of $\mu_D$. Thus, in addition to the scales $\mu_\pm$, we have a third scale $\mu_D$ such that the symbolic structure of the jet cross section is
\be
\sigma_{\rm jet}~=~{\rm PDF(\mu_-})\otimes |{\cal M}(\mu_-,\mu_+,\mu_D)|^2\otimes{\rm PDF}(\mu_+)\otimes D(z,\mu_D).
\label{eq:comp}
\ee
These three scales at LO (and another three at NLO) should be chosen to minimize the NLO (or NNLO) correction.

Formally, at fixed $\alpha_s$ order, the variation of each scale does not change the result. The advantage of choosing optimal scales is that part of the contribution (for example the ${\cal O}(\alpha_s^3)$ term in (\ref{eq:dg})) is then placed in the PDF where it will be {\it resummed} by DGLAP evolution. In this way we account for an important part of the higher-order contributions.

The three different scales  
provide the correct resummation in each PDF and in the $D$-function. On the other hand, this allows a better identification of the jet, and to reduce the probability of catching two different partons in the same jet cone. To achieve the latter objective it is better to work with a small jet cone size, $\Delta R$. On the other hand, this means that jet fragmentation should be described by a lower scale $\mu_D\sim p_T~\Delta R$, where $p_T$ is the jet transverse momentum. Simultaneously a reasonable scale in the PDFs is of the order of $p_T$. Moreover, for the Mercedes-like 3-jet configuration, corresponding to point-like production. the optimum expected scale is $\sim M_X$, where $M_X$ is the mass of the whole jet system (in analogy to the choice of scale for $Z$ boson production).  For back-to-back kinematics it is more natural to expect a scale $\sim p_T$.  Moreover for forward jet production in the `left' direction we do not have enough phase space for the evolution of the PDF$(\mu_-)$. Therefore we expect that a smaller value of $\mu_-$ will provide a better description of the process. That is, the final hierarchy of scales $\mu_D<\mu_-<\mu_+$ should provide the most convergent pQCD series.
In particular it looks reasonable to have $\mu_D\sim p_T~\Delta R$, ~$\mu_-\sim 0.5~p_T$ but decreasing for more forward jets and $\mu_+\sim p_T$ but increasing as the jet is more forward.

\subsection{Optimal choice of three scales at NLO}
Let us investigate this hierarchy in more detail, following the argumentation that led to (\ref{eq:i}). For simplicity, we work at NLO, though the procedure extends straightforwardly to NNLO and higher orders. 

Formally to calculate the optimal scales $\mu_+,~\mu_-,~\mu_D$ we have to consider the hard matrix element with off-mass-shell partons, and study its dependence on the virtuality of the incoming partons and outgoing jets. Another possibility would be to consider the known next order $\alpha_s$ correction. The corresponding formulae already include the respective dependences on the internal parton virtualities; that is, the NLO cross section accounts for the virtuality dependence of the LO matrix element.  The question is what is the best way to extract the contribution corresponding to the virtuality dependence of each individual parton from the full NLO cross section?

Recall, that DGLAP evolution is written in terms of collinear factorization. Therefore it is most convenient  to order the contributions in terms of {\it angles}.

To obtain the best LO description we consider the 2$\to$3 NLO subprocess.
In the centre-of-mass frame of the jets we first calculate the angles $\theta_i$ between the final parton with the {\it lowest} $p_T$ and the other four partons participating in the process\footnote{Note that $\theta$ is not the polar angle, but is the full angle between a pair of jets.}. DGLAP evolution produces configurations which are strongly ordered in angle.  Therefore it is natural to assign the contribution with the smallest angle $\theta_i$ to the evolution of parton $i$.  In other words we have taken the cross section for the NLO 2$\to$3 process and divided it into four parts $\sigma_i$ corresponding to the smallest $\theta_i$'s with $i=+,~ -,~ D,~4$.   In terms of (\ref{eq:i}), the factorization scale corresponding to the evolution of parton $i$ should be chosen to reproduce the value of $\sigma_i$.  Note that in our single jet inclusive cross section we do not consider the fragmentation of the second highest $p_T$ jet, $i=4$. Therefore the part of the cross section with the soft jet approximately collinear with jet 4 cannot be moved into the DGLAP evolution by any choice of the factorization scales.

\begin{figure} [h]
\begin{center}
\includegraphics[scale=0.65,angle=-0]{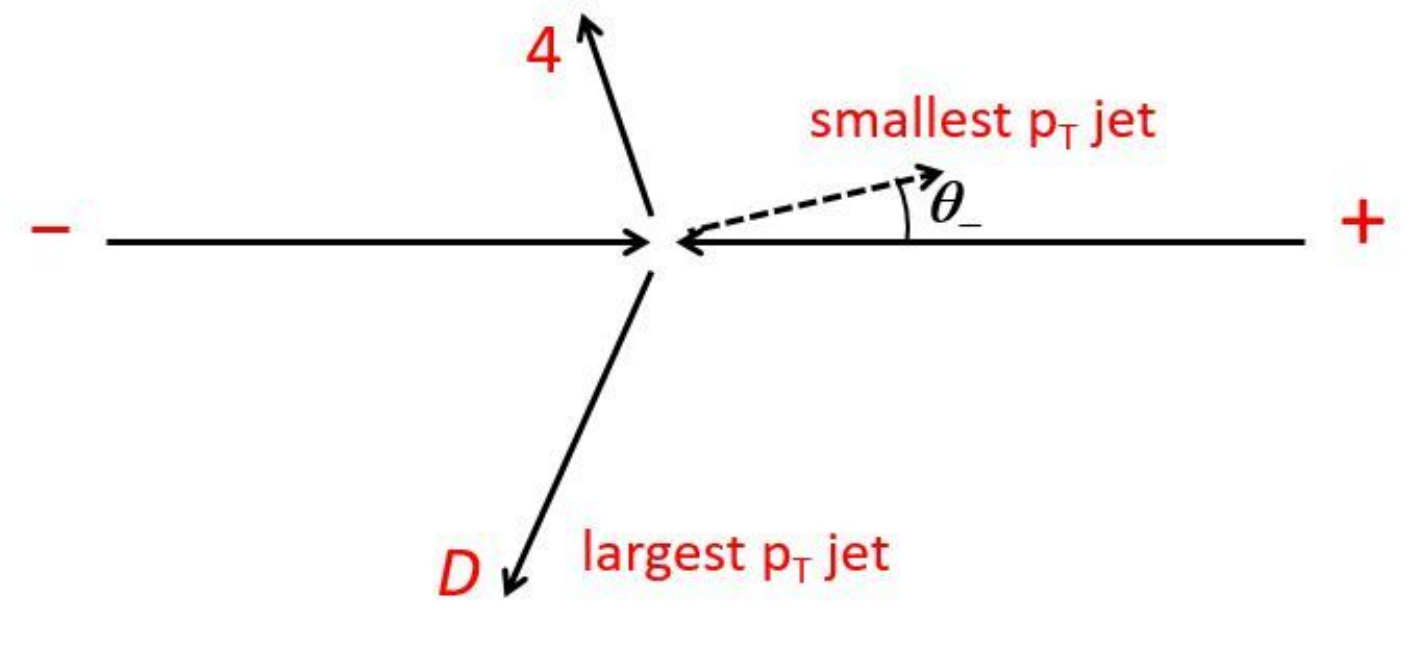}
\caption{\sf A configuration of three jet production in which the jet with the smallest transverse momentum $p_T$ aligns more closely with the incoming proton in the `$-$' direction.  For this part of the cross section, $\sigma_-$, the NLO emissions can be resummed and transferred to the LO PDF$(\mu_-)$ with optimal scale $\mu_-$. Similarly, those parts of the cross section where the smallest $p_T$ jet is aligned more closely with the incoming proton `$+$' direction or with the largest $p_T$ jet $D$ can be used to determine the optimal scales $\mu_+$ and $\mu_D$ respectively. }
\label{fig:f1}
\end{center}
\end{figure}

In summary, to determine the three scales for jet production at  at NLO, we divide up the 2$\to$3 cross section $\sigma$ into four parts, by measuring the angles to the smallest $p_T$ parton (jet), see Fig.~\ref{fig:f1}. To be more precise, we compute\footnote{In the case when the angle $\theta_4$ with respect to jet 4 is smaller than the jet cone size ($\Delta R>\theta_4$) and the sum of the transverse momenta $|\vec{p}_{T4}+\vec{p}_{T5}|>p_{TD}$ we have to consider jet 4 as the largest $p_T$ jet $D$.} 
\be
\sigma^{\rm NLO}_{j=+,-,D,4}~=~\sigma ~\Pi_{i\ne j}  ~ \Theta (\theta_i-\theta_j).
\ee
Then we choose scales $\mu_+,~\mu_-,~\mu_D$ such that the LO cross section calculated with these scales $\mu_j$ reproduces the corresponding part of $\sigma_{2\to 3}$ (like (\ref{eq:i})).  To do this it is convenient to start the calculation from some {\it low} dummy scale $\mu_0$. In this way we obtain a set of three equations, each of the form
\be
\sigma_j^{\rm NLO}(\mu_0)~=~|{\cal M}^{\rm LO}(k^2=0)|^2\otimes {\rm PDF}_{i\ne j}(\mu_0)\otimes D_{i\ne j}(\mu_0)\otimes {\rm PDF}_j(\mu_0)
\otimes P^{\rm real}(z)~{\rm ln}\frac{\mu_j^2}{\mu_0^2.}
\label{eq:j}
\ee
That is, we open the evolution of one of the PDF$_j$'s (or $D$) according to  (\ref{eq:i}). The NLO component of $\sigma_j$ describes the part of the evolution from $\mu_0$ to $\mu_j$ corresponding to the last term in (\ref{eq:i}). Note that since we deal with the $2\to 3$ subprocess we use only the component of the splitting function belonging to real emission.  In this way the set of three equations (\ref{eq:j})  determine the optimum values of $\mu_j$, with $j=+,-,D$, for the respective $j$ evolution. Since the loop corrections in DGLAP evolution are directly connected with the real emission component of the splitting function, in this way we also account for an important part of the NLO loop corrections. We explain how to avoid a possible soft gluon singularity in the Appendix.

To obtain more precise values of these $\mu_i$ we can perform a few iterations replacing for $i\ne j$ the dummy starting scale $\mu_0$ by $\mu_i$ from the previous iteration.

As is seen from the example given in (\ref{eq:stab1}), after the optimal scales are fixed, the final $\mu_f$ scale dependence of the predictions comes only from the variation of $\mu_f$ in the remaining NLO part of the cross section. This provides a much better factorization scale stability of the result. The same is true for the more general case of (\ref{eq:comp}).

A similar prescription may be applied to fix the scales in the NLO part of the cross section. There we select the NNLO contributions which are approximately collinear with the incoming partons or to the highest $p_T$ jet in the final state. These contributions can be moved and absorbed in the NLO PDFs and fragmentation function convoluted with the remaining NLO matrix element (see eq.(6) of \cite{OMR}).

Note that after the optimal factorization scales are fixed for the LO (NLO) part of the contribution, the dependence of the cross section on the universal scale (like $\mu_f$ in (\ref{eq:sta}) and (\ref{eq:stab1})) 
is considerably reduced, since it now comes only from the last term in (\ref{eq:stab1}) where the remaining coefficient function $C_{\rm rem}^{(1)} $ is small, while for (\ref{eq:sta}) both terms depend on $\mu_f$.

\subsection{Renormalization scale}

This paper concerns only the factorization scale dependence of the jet cross section.
Besides this, the pQCD prediction also depends on the renormalization scale, $\mu_R$.
Indeed, the NLO and NNLO expressions for jet cross section contain contributions up to ${\cal O}(\alpha_s^3(\mu_R))$ and ${\cal O}(\alpha_s^4(\mu_R))$ respectively. Recall that the choice of optimal factorization scales will, in general, reduce the higher $\alpha_s$ order contributions. Therefore we may expect that the dependence of the pQCD prediction on the renormalization scale, especially the part corresponding to configurations with three (or four) outgoing partons, will be reduced as well.

\subsection{$t\bar{t}$ production}

It is important to also find the optimal factorization scales for the $t\bar{t}$ differential cross sections. This process is dominantly driven by $gg$ fusion and allows an independent constraint on the large $x$ gluon PDF. Again data exist (see, for example, \cite{ATLAStop,CMStop}) and the NNLO formulation is known \cite{NNLOtop}. 

Exactly the same procedure can be used for the subprocesses $gg\to t{\bar t}g$ and $gg\to t{\bar t}gg$.
One first has to study the NLO 2$\to$3 (NNLO 2$\to$4) cross sections and to select the contributions $\sigma_j$ with the outgoing gluons (quarks) approximately collinear to the incoming gluons (quarks) $j$. These parts can be moved and resummed in the incoming  PDFs at the previous $\alpha_s$ order by choosing scales $\mu_\pm$ given by an equation analogous to (\ref{eq:j}).  As a rule the $t\bar{t}$ data are presented in terms of the $t$ quark. Therefore we have no problems with the fragmentation function\footnote{For example, in the $t\to b \mu \nu$ decay the momentum of the $b$ quark jet is close to that of the $B$ meson due to the strong leading effect. Therefore the effect of the resummation in the fragmentation function is minimal.}  and only two scales $\mu_\pm$ need to be optimized.

\section{The need for a Monte Carlo}

Since the jet is not an object that can be directly observed in a detector, it is usually defined as a group of secondaries emitted in a cone of size $\Delta R$. The precise inclusive jet cross section depends on the particular jet searching algorithm. In order to compare the experimental results with pQCD the experimentalists have to use a Monte Carlo event generator to account for the corrections caused by the detector efficiency, by hadronization, by experimental cuts and the effect of the underlying events. The problems are, first, that we have no NNLO Monte Carlo. Next, the present Monte Carlos do not have options to introduce different factorization scales in the three different components in (\ref{eq:comp}); that is, in the `left' and `right' PDFs and in the fragmentation function. Therefore K factors, which reflect the ratio of NNLO/NLO (or NNLO/LO) pQCD predictions, are used to correct the result obtained from the NLO (or LO) Monte Carlo.   Since the value of the NNLO remaining correction depends on the choice of factorization scale at the previous NLO (or LO) level, we have two possibilities.  Either to use the same unified scale as in the Monte Carlo, or, to obtain better precision, to calculate the NNLO+NLO+LO result using the different scales, as proposed\footnote{This will provide better accuracy in the numerator of the ratio.} in Section 3, but in calculating the NLO (or LO) denominator we still have to use the same universal scale as in the Monte Carlo.

There is no reason to expect the K factor to be the same for different kinematical configurations of the produced jets.
The higher $\alpha_s$ order corrections caused by the emission of additional jets clearly depend on the phase space available for one or another emission.

\section{Summary}
We emphasize the pQCD prediction for the cross section for inclusive high-$p_T$ jet production contains three different factorization scales.
The choice of these scales is an uncertainty in the description of the jet data by pQCD.
To improve convergence of the pQCD series we have shown that three different factorization scales may be used for the LO part  (and an additional three in NLO term and so on).  Two scales correspond to the incoming PDFs and the third to the jet fragmentation function, $D$.  Indeed, {\it all} the factorization scale dependence of the NLO (NNLO) matrix element (or coefficient function) comes from the subtraction of the contribution included in the PDF (or $D$) jet evolution. This subtraction is needed to avoid double counting of NLO (NNLO) contribution. We fix the factorization scale in each PDF (or $D$ function) to minimize the next $\alpha_s$ contribution. Then the part of the contribution transferred to the PDF (or $D$) is resummed to all $\alpha_s$ orders by the evolution. This provides a better pQCD description. 
 We have shown how to determine the {\it optimal} value of each factorization scale at NLO (NNLO) based on the knowledge of the NLO (NNLO) contribution and on the {\it collinear} nature of DGLAP evolution.  
 Having fixed the optimal scales in the lower $\alpha_s$ order term, the dependence of the cross section on the universal factorization scale $\mu_f$ is considerably reduced, since it now comes from the much smaller remaining higher $\alpha_s$ order term.

 Besides maximizing the convergence of the pQCD series, the introduction of different optimal scales also allows a better jet identification, since the scale used for jet evolution is causally connected with the jet cone size $\Delta R$. This allows the variation of $\Delta R$ without affecting the incoming PDFs.
 
 Finally we note that the proposed procedure can also be used to calculate the scales for inclusive $t\bar{t}$ production. In this case we need to optimize only two factorization scales.

\section*{Appendix: Absence of infrared contributions}
Here we explain why infrared contributions do not occur in the evaluation of $\sigma^{\rm NLO}_j$ of eq. (\ref{eq:j}).
First note that by starting with a small, but non-zero, scale $\mu_0$ we automatically avoid the 
infrared contribution coming from low parton virtuality, $k^2$. We are 
interested in the convergence of integral of (\ref{eq:i}) at the upper limit and 
study only the region of $k^2 > \mu^2_0$. The low $k^2<\mu^2_0$ domain is 
regularized in the usual way (appropriate for the NLO coefficient function and DGLAP 
evolution). 

Another possible problem is the `soft' singularity corresponding to 
the emission of a very soft gluon. That is, to the $1/(1-z)$ term in the 
splitting kernel. Theoretically this singularity is cancelled by the 
`self-energy' loop contribution and formally it is usually performed using 
the `plus' prescription\footnote{That is the integral $\int dz f(z)/(1-z)$ is 
replaced by $\int dz (f(z)-f(1))/(1-z)$.}.  In this case we  have to use the same `plus' prescription for the analytical calculations. 

However, it is worth mentioning what happens if a Monte Carlo (MC) were to be used to calculate the 
partial cross sections $\sigma^{\rm NLO}_j$ of (\ref{eq:j}).  Then we must
deal with the $1/(1-z)$ singularity on right-hand-side of (\ref{eq:j}) in exactly  
the same manner as that used in the MC. As a rule, in a MC, a cutoff in gluon 
transverse momenta, like $p_T>q_{\rm cut}$, is implemented. In such a case the 
 $z\to 1$ region corresponds to very high virtuality (or scale) 
$k^2=p^2_T/(1-z)$ and its contribution will be suppressed (regularised) by 
the LO matrix element $|{\cal M}^{\rm LO}(k^2)|^2$ which decreases with $k^2$.
 Anyway, just kinematically, we have the condition that the soft gluon 
energy $p_0>p_T$. This energy $p_0\propto (1-z)M_{jj}$ decreases faster as $z\to 
1$ than the values of $p_T>\sqrt{(1-z)}\mu_0$ for the process with 
scale $p^2_T/(1-z) > \mu^2_0$. Here $M_{jj}$ is the dijet mass.  This will introduce a natural cutoff
$(1-z)>(\mu_0/M_{jj})^2$. That is, actually in such MC calculations we will 
never face the singularity. However, in order not to sample an additional 
contribution from `soft' gluon emission we must take care to 
implement on the right-hand-side of (\ref{eq:j}) exactly the same `soft cutoff' as 
that used on the left-hand-side.

\section*{Acknowledgements}

We thank Nigel Glover and James Currie for interesting discussions and for stimulating our interest in this problem. We also thank Emmanuel de Oliveira for valuable comments. MGR thanks the IPPP at the University of 
Durham for hospitality. This work of MGR was supported by the RSCF grant 14-22-00281.

\end{document}